\documentstyle[12pt]{article}
\begin{document}

\def\p{\phi}
\def\P{\Phi}
\def\a{\alpha}
\def\e{\varepsilon}
\def\be{\begin{equation}}
\def\ee{\end{equation}}
\def\l{\label}
\def\0{\setcounter{equation}{0}}
\def\T{\hat{T}_}
\def\b{\beta}
\def\S{\Sigma}
\def\3{d^3{\rm \bf x}}
\def\4{d^4}
\def\C{\cite}
\def\r{\ref}
\def\ba{\begin{eqnarray}}
\def\ea{\end{eqnarray}}
\def\n{\nonumber}
\def\R{\right}
\def\L{\left}
\def\q{\hat{Q}_0}
\def\X{\Xi}
\def\x{\xi}
\def\la{\lambda}
\def\d{\delta}
\def\s{\sigma}
\def\f{\frac}
\def\vx{{\rm \bf x}}
\def\j{\frac{\delta}{i \delta j_a ({\rm \bf x},x_0+t+t_1)}}

\begin{titlepage}
\begin{flushright}
      {\normalsize IP GAS-HE-6/95}
\end{flushright}
\vskip 3cm
\begin{center}
{\Large\bf $S$-matrix description of the finite-temperature
noneqilibrium media}
\vskip 1cm

\mbox{J.Manjavidze}\footnote{Institute of Physics,
Georgian Academy of Sciences, Tamarashvili str. 6,
Tbilisi 380077, Republic of Georgia,
e-mail:~jm@physics.iberiapac.ge} \\

\end{center}
\date{MARCH 1995}
\vskip 1.5cm

\begin{abstract}
\footnotesize

The paper contains the real-time perturbation theory for description
of a statistical system with the nonuniform temperature distribution.
The formalism based on the  Wigner-functions approach. The perturbation
theory is formulated in terms of the  local-temperature Green functions.

\end{abstract}
\end{titlepage}

\section{Introduction}
\setcounter{equation}{0}

The aim of this article is to construct the perturbation theory for
generating functional of Wigner functions \C {1,2,3} for the case of
nonuniform temperature distribution. As an example  of interesting
system one can have in mind  the process of a very large number of
hadrons creation at the high-energy collisions.
The phenomenology of high-multiplicity processes was given in \C {4,M}.

In terms of QCD
it is the rear  process of the cold quark-gluon plasma formation.
Qualitatively this is a process of total dissipation of high
kinetic energy density initial state. One can consider this process
also as the process of dissipation of a high-temperature local
fluctuation in a low temperature equilibrium media. Generally speaking, the
temperature  freely evolves in this processes and is not distributed
uniformly at least on the early stages.
In this  paper we will consider the general theory of such processes
which can be used not only in the particles physics.

We should adopt the standard $S$-matrix formalism which is applicable
to any nonequilibrium processes. In this microcanonical approach the
temperature $T$ will be introduced as  the Lagrange multiplier and
the physical (measurable) value of $T$ will be defined by the  equation of
states at the very end of calculations. Using standard terminology
\C{5}, we  will deal with the ``mechanical" perturbations only \C{6}
and it will not be necessary to divide the perturbations on ``thermal"
and ``mechanical" ones \C{7,8} (see also Sec.5).

The usual Kubo-Martin-Schvinger (KMS) periodic boundary conditions
\C {6,9} can not be applied here since they are applicable  for the
equilibrium case only \C{10} (see also \C{12}).
We will introduce the boundary conditions ``by hands", modeling the
environment of the system \C{12}. Supposing that the system is in a
vacuum we will have usual field-theoretical vacuum boundary
condition (Sec.2). We will consider also the system in the
background field of black-body radiation (Sec.4). Last one restores
the theory with KMS  boundary condition in the equilibrium limit.

Calculating the generating functional of Wigner functions the
local temperature distribution will be introduced:
$T(\vec{x},t)=1/\beta(\vec x ,t)$
is the temperature in the measurement point $(\vec{x},t)$ (Sec.3). In
other words, we will divide the ``measuring device" (but not the
system as usually was done, \C {13}) on the cells the dimension of
which tends to zero. The differential measure $D\beta(\vec{x},t)$ will
be defined taking into account the energy-momentum conservation law.

\section{Vacuum boundary condition}
\setcounter{equation}{0}

The probability $r(P)$ of in- into out-states transition with fixed total
4-momentum $P$ can be calculated using the $n$- into $m$-particle transition
amplitude $a_{n,m}$:
\ba
r(P)=\sum_{n,m}\frac{1}{n!m!}\int d\omega_{n}(q)\omega_{m}(p)\times
\n\\ \times
\delta^{(4)}(P-\sum^{n}_{k=1}q_{k})
\delta^{(4)}(P-\sum^{m}_{k=1}p_{k})
|a_{n,m}|^2,
\l{2.1}
\ea
where
\be
d\omega_{n}(q)=\prod^{n}_{k=1}d\omega (q_k)=
\prod^{n}_{k=1}\frac{d^{3}q_k}{(2\pi)^3 2\epsilon (q_k)}, \;
\epsilon (q)=(q^2 +m^2)^{1/2}.
\l{2.2}
\ee
Eq. (\ref{2.1}) is the basic formula of our calculations. The microcanonical
description was introduced in \C {12} considering the Fourier transformation
of $\d$-functions.

The amplitude $a_{n.m}$ looks as follows \C {12}:
\be
a_{n,m}((q)_n,(p)_m)=\prod^{n}_{k=1}\hat{\phi}(q_k)
\prod^{m}_{k=1}\hat{\phi}^* (p_k) Z(\phi),
\l{2.3}
\ee
where $q_k (p_k )$ are the momentum of in(out)-going particles and the
annihilation operator
\be
\hat{\phi}(q)=\int d^4 x e^{-iqx} \hat{\phi}(x), \;
\hat{\phi}=\frac{\delta}{\delta\phi (x)},
\l{2.4}
\ee
was introduced. Correspondingly, $\hat\p^*(p)$ is the creation operator.
One can put the  auxiliary field $\p(x)$ equal to zero at the end
of calculation. The vacuum into vacuum transition amplitude in presence of
external field $\phi$
\be
Z(\phi)=\int D\Phi e^{iS_{C_{+}}(\Phi)-iV_{C_{+}}(\Phi +\phi)}
\l{2.5}
\ee
is defined on the Mills' complex time contour $C_+$ \C {14}, i.e.
$C_+:t \rightarrow t+i\e ,\e >0$. In eq.(\ref{2.5}) $S_{C_+}$ is
the free part of the action and $V_{C_+}$ describes the interactions.

In this section we will propose the vacuum boundary condition:
\be
\int_{\sigma_{\infty}} d\sigma_{\mu}\Phi \partial^{\mu}\Phi =0
\l{2.6}
\ee
where $\s_{\infty}$ is the infinitely far hypersurface.

We  start consideration from the assumption that the
temperature fluctuations are large scale. In a cell the
dimension of which is much smaller then the fluctuation scale of
temperature we can assume that the temperature is a ``good"
parameter. (The ``good" parameter means that the corresponding fluctuations
are Gaussian.)

Let us surround the  interaction region, i.e. the system under consideration,
by $N$ cells with known space-time position and let us
propose that we can measure the energy and momentum of groups of
in- and out-going particles in each cell. The 4-dimension of cells
can not be arbitrary small in this case because
of the quantum uncertainty principle.

To describe this situation we decompose $\d$-functions in (\ref{2.1})
on the product of $(N+1)$ $\d$-functions:
\be
\delta^{(4)}(P-\sum^{n}_{k=1}q_k)=
\int
\prod^{N}_{\nu =1}\{dQ_{\nu}\delta (Q_{\nu}-\sum^{n_{\nu}}_{k=1}q_{k,\nu})\}
\delta^{(4)}(P-\sum^{N}_{\nu =1}Q_{\nu}),
\l{2.7}
\ee
where $q_{k,\nu}$ are the momentum of $k$-th in-going particle in the
$\nu$-th cell and $Q_{\nu}$ is the total 4-momenta of $n_{\nu}$ in-going
particles in this cell. The same decomposition will be used for the
second $\d$-function in (\ref{2.1}). Inserting this decompositions into
(\ref{2.1}) we must take into account the multinomial character of
particles decomposition on $N$ groups. This will give the coefficient:
\be
\frac{n!}{n_{1}!\cdots n_{N}!}\delta_{K}(n-\sum^{N}_{\nu =1}n_{\nu})
\frac{m!}{m_{1}!\cdots m_{N}!}\delta_{K}(m-\sum^{N}_{\nu =1}m_{\nu}),
\l{2.8}
\ee
where $\d_{K}$ is the Kronecker's $\d$-function.

In result, the quantity
\ba
r((Q)_N,(P)_N)=
\sum_{(n.m)} \int |a_{(n,m)}|^2\times
\n\\ \times
\prod^{N}_{\nu =1}\{ \prod^{n_{\nu}}_{k=1}\frac{d\omega (q_{k,\nu})}{n_{\nu}!}
\delta^{(4)}(Q_{\nu}-\sum^{n_{\nu}}_{k=1}q_{k,\nu})
\prod^{m_{\nu}}_{k=1}\frac{d\omega(p_{k,\nu})}{m_{\nu}!}
\delta^{(4)}(P_{\nu}-\sum^{m_{\nu}}_{k=1}p_{k,\nu})\}
\l{2.9}
\ea
describes  the probability that in the $\nu$-th cell we measure the fluxes
of in-going particles with total 4-momentum  $Q_{\nu}$ and of out-going
particles with the total 4-momentum $P_{\nu}$. The sequence of this two
measurements is not fixed.

The Fourier transformation of $\d$-functions in (\ref{2.9}) gives the
formula:
\be
r((Q)_N,(P)_N)=\int \prod^{N}_{k=1}
\frac{d^4 \alpha_{-,\nu}}{(2\pi)^4}
\frac{d^4 \alpha_{+,\nu}}{(2\pi)^4}
e^{i\sum^{N}_{\nu =1}(Q_{\nu}\alpha_{-,\nu} +
P_{\nu}\alpha_{+,\nu})}
R((\alpha_-)_N,(\alpha_+)_N),
\l{2.10}
\ee
where $R((\alpha_- )_N,(\alpha_+ )_N )=R(\alpha_{-,1},
\alpha_{-,2}...,\alpha_{-,N};
\alpha_{+,1},\alpha_{+,2},...,\alpha_{+,N})$
has the form:
\ba
R((\alpha_-)_N,(\alpha_+)_N)
=\int
\prod_{\nu =1}^{N}\{\prod^{n_{\nu}}_{k=1} \frac{d\omega(q_{k,\nu})}{n_{\nu}!}
e^{-i\alpha_{-,\nu}q_{k,\nu}}\times
\n \\ \times
\prod^{m_{\nu}}_{k=1} \frac{d\omega(p_{k,\nu})}{m_{\nu}!}
e^{-i\alpha_{+,\nu}p_{k.\nu}}\}
|a_{(n,m)}|^2.
\l{2.11}
\ea
Inserting (\ref{2.3}) into (\ref{2.11}) we find:
\ba
R((\alpha_-)_N,(\alpha_+)_N)=
\exp\{ i\sum_{\nu =1}^{N} \int dx dx' [
\hat{\phi}_+ (x)D_{+-}(x-x';\alpha_{+,\nu})\hat{\phi}_- (x')-
\n\\
-\hat{\phi}_- (x)D_{-+}(x-x';\alpha_{-,\nu})\hat{\phi}_+ (x')]\}
Z(\phi_+)Z^*(\phi_-),
\l{2.12}
\ea
where $\p_-$ is  defined on the complex conjugate contour
$C_-:t\rightarrow t-i\e$ and
\be
D_{+-}(x-x';\alpha)=-i\int d\omega (q) e^{iq(x-x')} e^{-i\alpha q},
\l{2.13}
\ee

\be
D_{-+}(x-x';\alpha )=i\int d\omega (q) e^{-iq(x-x')} e^{-i\alpha q}
\l{2.14}
\ee
are the positive and negative frequency correlation
functions correspondingly.

We must integrate over sets $(Q)_N$ and $(P)_N$ if the
distribution of fluxes momenta over cells is not fixed. In result,
\be
r(P)=\int D^{4}\alpha_- (P) d^{4}\alpha_+ (P)
R((\alpha_-)_N ,(\alpha_+)_N),
\l{2.15}
\ee
where the differential measure
\be
D^{4}\alpha (P)=\prod^{N}_{\nu =1} \frac{d^4 \alpha_{\nu}}{(2\pi)^4}
K(P,(\alpha)_N)
\l{2.16}
\ee
takes into account the energy-momentum conservation laws:
\be
K(P,(\alpha)_N)=
\int \prod^{N}_{\nu =1} d^4 Q_{\nu}
e^{i\sum^{N}_{\nu =1}\alpha_{\nu}Q_{\nu}}
\delta^{(4)}(P-\sum^{N}_{\nu =1}Q_{\nu}).
\l{2,17}
\ee
The explicit integration gives that
\be
K(P,(\alpha)_N)\sim \prod^{N}_{\nu =1} \delta^{(3)}(\alpha -\alpha_{\nu}),
\l{2.18}
\ee
where $\vec{\alpha}$ is the center of mass (CM) 3-vector.

To simplify the consideration let us choose the CM frame and put
$\alpha=(-i\beta ,\vec{0})$. In result,
\be
K(E,(\beta)_N)=\int^{\infty}_{0} \prod^{N}_{\nu =1} dE_{\nu}
e^{\sum^{N}_{\nu =1}\beta_{\nu}E_{\nu}}
\delta(E-\sum^{N}_{\nu =1} E_{\nu})
\l{2.19}
\ee
Correspondingly, in the CM frame,
\be
r(E)=\int D \beta_+ (E) D \beta_- (E) R((\beta_+ )_N,(\beta_- )_N),
\l{2.20}
\ee
where
\be
D \beta (E)=\prod^{N}_{\nu=1}\frac{d \beta_{\nu}}{2\pi i}K(E,(\beta)_N)
\l{2.21}
\ee
and $R((\beta)_N)$ was defined in (\ref{2.12})
with $\alpha_{k,\nu}=(-i\beta_{k,\nu},\vec0),\;\;
Re\beta_{k,\nu} >0,\;\;k=+,-$.

We will calculate integrals  over $\beta_k$ using the stationary phase
method. The equations for mostly probable values of $\beta_k$:
\be
-\frac{1}{K(E,(\beta_k)_N)}\frac{\partial}{\partial \beta_{k,\nu}}
K(E,(\beta_k)_N)=
\frac{1}{R((\beta_1)_N)}
\frac{\partial}{\partial \beta_{k,\nu}}
R((\beta)_N),\;\;\;k=+,-,
\l{2.22}
\ee
always has the unique positive solutions $\tilde{\beta}_{k,\nu}(E)$. We
propose that the fluctuations of $\beta_{k}$ near $\tilde{\beta}_k$
are small, i.e. are Gaussian. This is the basis of the local-equilibrium
hypothesis \C {13}. In this case $1/\tilde{\beta}_{-,\nu}$ is the temperature
in the initial state in the measurement cell $\nu$ and
$1/\tilde{\beta}_{+,\nu}$ is the temperature of the final state  in the
$\nu$-th measurement cell.

The last formulation (\ref{2.15}) imply that the 4-momenta $(Q)_N$ and
$(P)_N$ can not be measured. It is possible to consider another
formulation also. For instance, we can suppose that the initial set
$(Q)_N$ is fixed (measured) but $(P)_N$ is not. In this case we
will have mixed experiment: $\tilde{\beta}_{-,\nu}$ is defined by the
equation:
\be
E_{\nu}=-\frac{1}{R}
\frac{\partial}{\partial \beta_{-,\nu}}R
\l{2.23}
\ee
and $\tilde{\beta}_{+,\nu}$ is defined by second equation in (\ref{2.22}).

Considering limit $N\rightarrow \infty$ the dimension of cells
tends to zero. In this case we are forced by quantum uncertainty
principle to propose that the 4-momenta sets $(Q)$ and
$(P)$ are not fixed. This formulation becomes pure thermodynamical:
we must assume that $(\b_-)$ and $(\b_+)$ are measurable quantities.
For instance, we can fix $(\b_-)$ and try to find $(\b_+)$ as the
function of total energy $E$ and the functional of $(\b_-)$.
In this case eqs.(\ref{2.22}) become the functional equations.

In the  considered microcanonical description the finiteness of
temperature does not touch the quantization mechanism. Really, one
can see from (\ref{2.12}) that all thermodynamical information is
confined in the operator exponent
\be
e^{N(\hat{\phi}_i\hat{\phi}_j)}=
\prod_{\nu}\prod_{i\neq j}e^{i\int \hat{\phi}_{i} D_{ij}\hat{\phi}_{j}}
\l{2.24}
\ee
the expansion of which describes the environment, and the ``mechanical"
perturbations are  described by the amplitude $Z(\phi)$. This
factorization was achieved by introduction of auxiliary field $\p$ and is
independent from the choice of boundary conditions, i.e. from the choice
of the considered systems environment.

\section{The distribution functions}
\setcounter{equation}{0}

In the previous  section the generating functional $R((\beta)_N)$
was calculated by means of dividing the ``measuring device"
(calorimeter) on the $N$ cells. It was assumed that the dimension
of device cells tends to zero ($N\rightarrow \infty$). Now we
will specify the cells coordinates using the Wigner's description
\C {1,2,3}.

Let us introduce the distribution function $F_n$ which defines the
probability to find $n$ particles with definite momentum and with
arbitrary coordinates. This probabilities (cross section) are
usually measured in particle physics. The corresponding Fourier-transformed
generating functional can be deduced from (\ref{2.12}):
\ba
F(z,(\beta_+)_N ,(\beta_-)_N)=
\prod^{N}_{\nu =1}
\prod_{i\neq j}e^{\int d\omega (q)
\hat{\phi}^*_i (q)
e^{-\beta_{j,\nu}\epsilon (q)}
\hat{\phi}_j (q) z^{\nu}_{ij}(q)}\times
\n \\\times
Z(\phi_+)Z^*(\phi_-).
\l{3.1}
\ea
The variation of $F$ over $z^{\nu}_{ij}(q)$ generates corresponding
distribution functions. One can interpret $z^{\nu}_{ij}(q)$ as the
local activity: the logarithm of $z^{\nu}_{ij}(q)$ is conjugate to
the particles number in the cell $\nu$ with momentum $q$ for the
initial ($ij=+-$) or final ($ij=-+$) states. Note that
$z^{\nu}_{ij}(q)\hat{\phi}^*_i (q)\hat{\phi}_j (q)$ can be
considered as the operator of activity.

The Boltzman factor $e^{-\beta_{i,\nu} \epsilon (q)}$ can be interpreted as
the probability to find a particle with the energy $\epsilon (q)$ in the
final state ($i=+$) and in the initial state ($i=-$).
The total probability, i.e. the process of creation and
further absorption of $n$ particles, is defined by multiplication of
this  factors.

The generating functional (\ref{3.1}) is  normalized as follows:
\be
F(z=1,(\beta ))=R((\beta)),
\l{3.2}
\ee
\be
F(z=0,(\beta))=|Z(0)|^2=R_0 (\phi_{\pm})|_{\phi_{\pm}=0}
\l{3.3}
\ee
Where
\be
R_0 (\phi_{\pm})=
Z(\phi_+)Z^*(\phi_-)
\l{3.4}
\ee
is the ``probability" of the vacuum into vacuum transition in presence
of auxiliary fields $\phi_{\pm}$. The one-particle distribution function
\ba
F_1 ((\beta_+)_N,(\beta_-)_N;q)
=\frac{\d}{\d z^{\nu}_{ij}(q)} F|_{z=0}=
\n \\
=\{\hat{\phi}^*_i (q)e^{-\beta^{\nu}_{i}\epsilon (q)/2}\}
\{\hat{\phi}_j (q)e^{-\beta^{\nu}_{i}\epsilon (q)/2}\}
R_0 (\phi_{\pm})
\l{3.5}
\ea
describes the probability to find one particle in the vacuum.

Using definition (\ref{2.4}),
\ba
F_1 ((\beta_+)_N,(\beta_-)_N;q)=
\int dx dx' e^{iq(x-x')}e^{-\beta_{i,\nu}\epsilon (q)}\}
\hat{\phi}_i (x)\hat{\phi}_j (x')R_0(\phi_{\pm})=
\n \\
=\int dY \{dy e^{iqy}e^{-\beta_{i,\nu}\epsilon (q)}\}
\hat{\phi}_i (Y+y/2)\hat{\phi}_j (Y-y/2)R_0(\phi_{\pm})\}.
\l{3.6}
\ea
We introduce using this definition the one-particle Wigner function $W_1$
\C {2}:
\be
F_1 ((\beta_+)_N,(\beta_-)_N;q)=
=\int dY W_1 ((\beta_+)_N,(\beta_-)_N;Y,q).
\l{3.7}
\ee
So,
\be
W_1 ((\beta_+)_N,(\beta_-)_N;Y,q)=\int dy e^{iqy}e^{-\beta_{i,\nu}\e (q)}
\hat{\phi}_i (Y+y/2)\hat{\phi}_j (Y-y/2)R_0(\phi_{\pm}).
\l{3.8}
\ee
This distribution function describes the probability to find
in the vacuum particle with momentum $q$ at the point $Y$ in the
cell $\nu$

Since the choice of the device coordinates is in our hands it is
natural to adjust the cell coordinate $\n$ to the
coordinate of measurement $Y$:
\be
W_1((\beta_+)_N,(\beta_-)_N;Y,q)=\int dy
e^{iqy}e^{-\beta_{i}(Y)\epsilon (q)}\}
\hat{\phi}_i (Y+y/2)\hat{\phi}_j (Y-y/2)R_0(\phi_{\pm}).
\l{3.9}
\ee
This choice of the device coordinates lead to the following
generating functional:
\ba
F(z,\beta)=
\exp\{i\int dydY [\hat{\phi}_+ (Y+y/2)D_{+-}(y;\beta_+ (Y),z)
\hat{\phi}_- (Y-y/2)-
\n \\
-\hat{\phi}_- (Y+y/2)D_{-+}(y;\beta_- (Y),z)\hat{\phi}_+ (Y-y/2)]\}
R_0(\phi_{\pm}),
\l{3.10}
\ea
where
\be
D_{+-}(y;\beta_+ (Y),z)=-i\int d\omega (q) z_{+-}(Y,q)
e^{iqy}e^{-\beta_{+}(Y)\epsilon (q)},
\l{3.11}
\ee
\be
D_{-+}(y;\beta_+ (Y),z)=i\int d\omega (q) z_{-+}(Y,q)
e^{-iqy}e^{-\beta_{-}(Y)\epsilon (q)}
\l{3.12}
\ee
are the modified positive and negative correlation functions
(\ref{2.13}), (\ref{2.14}).

The inclusive, partial, distribution functions are familiar
in the particle physics. This functions describe
the distributions in presence of arbitrary number of other particles.
For instance, one-particle partial distribution function
\ba
P_{ij} (Y,q;(\beta ))
=\frac{\d}{\d z_{ij}(Y,q)} F(z,(\beta))|_{z=1}=
\n \\
=\frac{e^{-\beta_{i}(Y)\epsilon (q)}}
{(2\pi)^3 \epsilon (q)}\int dy e^{iqy}
\hat{\phi}_i (Y+y/2)\hat{\phi}_j (Y-y/2)
R(\phi_{\pm},(\beta)),
\l{3.13}
\ea
where eq.(\ref{3.2}) was used.

The mean multiplicity
$n_{ij}(Y,q)$ of particles in the infinitesimal cell $Y$ with
momentum $q$ is
\be
n_{ij}(Y,q)
=\int dq \frac{\d}{\d z_{ij}(Y,q)} \ln F(z,(\beta))|_{z=1}.
\l{3.14}
\ee
If the interactions among fields are switched out we can find that
(omitting indexes):
\be
n(Y,q_0)=\frac{1}{e^{\beta (Y)q_0}-1},\;\;q_0=\epsilon (q)>0.
\l{3.15}
\ee
This is the mean multiplicity of black-body radiation.

\section{The closed-path boundary condition}
\setcounter{equation}{0}

The developed in Sec.2 formalism allows to introduce the more general
boundary conditions instead of (\ref{2.5}). Considering the probability
$R$ which has the double path integral representation we will introduce
integration over closed path. This allows to introduce the equality:
\be
\int_{\sigma_{\infty}} d\sigma_{\mu}
(\Phi_+ \partial^{\mu}\Phi_+ -
\Phi_- \partial^{\mu}\Phi_-)=0,
\l{4.1}
\ee
as  the boundary condition, where $\sigma_{\infty}$ is the infinitely
far hypersurface. The general solution of this equation is:
\be
\Phi_{\pm}(\sigma_{\infty})=
\Phi (\sigma_{\infty})
\l{4.2}
\ee
where $\Phi(\sigma_{\infty})$ is the ``turning-point" field.
The result of this changing of boundary condition was analyzed in
\C {12} for the case of uniform temperature distribution.

In terms  of $S$-matrix the field $\Phi(\sigma_{\infty})$
represent the background flow of mass-shell particles.
We will propose that the probability to find a particle of
the background flow is determined by the energy-momentum
conservation law only. In another words, we  will propose
that the system under consideration is surrounded by the
black-body radiation.

Presence of additional flow will reorganize the differential
operator $\exp\{N(\hat{\phi}_i \hat{\phi}_j)\}$ only and new
generating functional $R_{cp}$ has the form:
\be
R_{cp}(\alpha_+,\alpha_-)=e^{N(\hat{\phi}_i\hat{\phi}_j)}
R_0 (\phi_{\pm}).
\l{4.3}
\ee
The calculation of operator $N(\hat{\phi}_i \hat{\phi}_j)$
is strictly the same as in \C {12}. Introducing the
cells in the $Y$ space we will find that
\be
N(\hat{\phi}_i\hat{\phi}_j)=
\int dY dy \hat{\phi}_i(Y+y/2) \tilde{n}_{ij}(Y,y)\hat{\phi}_j(Y-y/2),
\l{4.4}
\ee
where the occupation number $\tilde{n}_{ij}$ carries the cells index
$Y$:
\be
\tilde{n}_{ij}(Y,y)=\int d\omega (q) e^{iqy}n_{ij}(Y,q)
\l{4.5}
\ee
and ($q_0 =\epsilon (q)$)
\be
n_{++} (Y,q_0)=n_{--}(Y,q_0)=\tilde{n}(Y,(\beta_+ +\beta_-)|q_0|/2)=
\frac{1}{e^{(\beta_+ +\beta_-)(Y)|q_0|/2}-1},
\l{4.6}
\ee
\be
n_{+-}(Y,q_0)=\Theta (q_0)(1+\tilde{n}(Y,\beta_+ q_0))+
\Theta (-q_0)\tilde{n}(Y,-\beta_- q_0),
\l{4.7}
\ee
\be
n_{-+}(Y,q_0)=n_{+-}(Y,-q_0).
\l{4.7'}
\ee
For simplicity the CM system was used.

Calculating $R_0$ perturbatively we will find that
\ba
R_{cp}(\beta)=\exp\{-iV(-i\hat{j}_+)+iV(-i\hat{j}_-)\}\times
\n \\\times
\exp\{i\int dY dy[\hat{j}_i (Y+y/2)G_{ij}(y,(\beta (Y))\hat{j}_j (Y-y/2)\}
\l{4.8}
\ea
where, using the matrix notations,
\ba
iG(q,(\beta (Y)))=
\left(
\matrix{
\frac{i}{q^2 -m^2 +i\e } & 0 \cr
0 & -\frac{i}{q^2 -m^2 -i\e } \cr
}\right)
+\n \\ \n \\+
2\pi \d (q^2 -m^2)
\left(
\matrix{
n(\frac{(\beta_+ +\beta_-)(Y)}{2}|q_0|) &
n(\beta_+ (Y)|q_0|)a_+ (\beta_+) \cr
n(\beta_-(Y)|q_0|)a_- (\beta_-) &
n(\frac{(\beta_+ +\beta_- )(Y)}{2}|q_0|) \cr
}\right),
\l{4.9}
\ea
and
\be
g_{\pm}(\beta)=-e^{\beta (|q_0|\pm q_0)/2}.
\l{4.9'}
\ee
Formally this Green functions obey the standard equations in the
$y$ space:
\ba
(\partial^2 -m^2)_y G_{ii}=\delta (y),
\n \\
(\partial^2 -m^2)_y G_{ij}=0, \;\;i \neq j
\l{4.10}
\ea
since $\Phi (\s_{\infty}) \neq 0$ reflects the mass-shell particles.
But the boundary conditions for this equations are not evident.

\section{Concluding remarks}
\setcounter{equation}{0}

One  can not expect the evident connection between the above considered
and Zubarev's \C{13} approaches. The reason is as follows.

In Zubarev's theory the ``local-equilibrium" hypothesis was adopted
as the boundary condition.It is assumed that in the suitably
defined cells of a system at a given temperature distribution
$T(\vec{x},t)=1/\beta(\vec x,t)$ where $(\vec x ,t)$ is the index of
the cell, the entropy is maximum. The corresponding nonequilibrium
statistical operator
\be
R_z \sim e^{-\int d^{3}x \beta T_{00}}
\l{1.1}
\ee
describes  evolution of a system. Here $T_{\mu \nu}$ is the
energy-momentum tensor. It is assumed that the system
``follows" to $\beta(\vec x, t)$ evolution and the local temperature
$T(\vec{x},t)$is defined as the external parameter which is
the regulator of systems dynamics. For this purpose the special
$i\e$-prescription was introduced \C{13}.

The KMS periodic boundary condition \C {5,9} can not be applied
\C {10,12} and by this reason the decomposition:
\be
\beta (\vec{x},t)=\beta_{0}+\beta_{1}(\vec{x},t)
\l{1.2}
\ee
was offered in the paper \C {7}. Here $\beta_0$ is the constant and
the inequality
\be
\b_{0}>>|\beta_{1}(\vec{x},t)|
\l{1.3}
\ee
is  assumed. Then,
\be
R_{z}\sim e^{-\beta_{0}(H_{0}+V+B)}
\l{1.4}
\ee
where $H_0$ is the  free part of the  Hamiltonian, $V$ describes
the interactions and the linear over $\beta_1 /\beta_0$ term $B$ is
connected with the deviation of temperature from the ``equilibrium"
value $1/\beta_0$. Considering $V$ and  $B$ as the perturbations
one can calculate the observables averaging over equilibrium
states, i.e. adopting the KMS boundary condition. Using standard terminology
\C {5} one  can consider $V$ as the ``mechanical" and $T$ as the
``thermal" perturbations.

The quantization problem of operator  (\ref{1.4}) is connected
with definition of the space-time sequence of mechanical ($V$)
and thermal ($B$) excitations. It is necessary since the mechanical
excitations give the influence on the thermal ones and vice versa.
It was assumed in \C {7} that $V$ and $B$ are commuting operators,
i.e. the sequence of $V$- and $B$-perturbations is not sufficient.
This solution leads to the particles propagators renormalization by
the interactions with the external field $\beta(\vec x, t)$ even without
interactions among fundamental fields. (Note absence of this
renormalizations in our formalism.)

In \C {8} the operators $V$ and $B$ are  noncommuting ones and
$B$-perturbations were switched on after $V$-perturbations.
In this formulation the nondynamical renormalization are also
present but it is not unlikely that they are
canceled at the very end of calculations \C {15}.

This formulation with $\beta(\vec x, t)$ as the external field
remained the old, firstly quantized, field theory in which matter
is quantized but fields are not. It is known that consistent quantum
field theory requires the second quantization. Following to this
analogy, if we want to take into account consistently the reciprocal
influence of $V$- and $B$-perturbations the field $\beta(\vec{x}, t)$
must be fundamental, i.e. must be quantized (and the assumption of
paper \C {7} becomes true). But it is evidently the wrong idea in
the canonical Gibbs formalism. So, as in the firstly quantized theory,
the theory with operator (\ref{1.1}) must have the restricted
range of validity \C{13}.

Therefore, we must reduce our formalism just to the hydrodynamical
accuracy to find the connection with Zubarev's approach. There
is the another side of this question. The offered formalism is able
to describe an arbitrary nonequilibrium process since it based on
the $S$-matrix, i.e. on the strict field-theoretical description.
But the mechanism of irreversibility is not clearly seen:
the generating functional $R_0 (\phi_{\pm})$ is described by
the  closed-path motion in the  functional space, i.e. formally is the
time-reversible quantity.

\vspace {0.2in}
{\Large \bf Acknowledgement}
\vspace {0.2in}

I would like to thank T.Bibilashvili for interesting discussions.
This work was supported in part by the  U.S.~National Science
Foundation.

\end{document}